\documentclass[fleqn,usenatbib]{mnras}
\usepackage{newtxtext,newtxmath}
\usepackage[T1]{fontenc}
\usepackage{ae,aecompl}
\DeclareRobustCommand{\VAN}[3]{#2}
\let\VANthebibliography\thebibliography
\def\thebibliography{\DeclareRobustCommand{\VAN}[3]{##3}\VANthebibliography}
\usepackage[flushleft]{threeparttable}
\usepackage{graphicx}	
\graphicspath{ {./figures/} }
\usepackage{amsmath}	
\usepackage{tabularx}
\usepackage{makecell, multirow}
\usepackage{gensymb}

\usepackage{soul}

\usepackage{upgreek}
\usepackage{float}

\usepackage{booktabs}

\title[Multi-wavelength study of the Blazar 4C\,+01.02]{Multi-wavelength study of blazar 4C\,+01.02 during its long-term flaring activity in 2014-2017}

\author[Malik Zahoor et al.]{
Malik Zahoor$^{1}$\thanks{E-mail: malikzahoor313@gmail.com},
Shah Zahir$^{2}$\thanks{E-mail: shahzahir4@gmail.com},
Sunder Sahayanathan$^{3,4}$\thanks{E-mail: sunder@barc.gov.in},
Naseer Iqbal$^{1}$,
Aaqib Manzoor$^{1}$
\\
$^{1}$Department of Physics, University of Kashmir, Srinagar 190006, India.\\
$^{2}$Department of Physics, Central University of Kashmir, Ganderbal 191201, India.\\
$^{3}$Astrophysical Sciences Division, Bhabha Atomic Research Center, Mumbai 400085, India.\\
$^{4}$Homi Bhabha National Institute, Mumbai 400094, India.
 }

\date{Accepted XXX. Received YYY; in original form ZZZ}

\pubyear{2021}

\begin{document}
\label{firstpage}
\pagerange{\pageref{firstpage}--\pageref{lastpage}}
\maketitle

\begin{abstract}
We conducted a detailed long-term spectral and temporal study of flat spectrum radio quasar 4C\,+01.02, 
by using the multi-wavelength observations from \emph{Fermi}-LAT, \emph{Swift}-XRT,  and \emph{Swift}-UVOT.
The $2$-day bin $\gamma$-ray lightcurve in the 2014-2017 active state displays $14$ peak structures with a maximum integral flux $(\rm E > 100 \ MeV)$ of $\rm (2.5 \pm 0.2)  \times 10^{-6}\ ph\ cm^{-2}\ s^{-1}$ at MJD 57579.1, which is approximately $61$ times higher than the  base flux of $\rm (4.1 \pm 0.3) \times 10^{-8}\ ph\ cm^{-2}\ s^{-1}$, calculated by averaging the flux points when the source was in quiescent state. The shortest $\gamma$-ray variability of $0.66 \pm 0.08$ days is observed for the source.
The correlation study between $\gamma$-ray spectral index and flux suggests that the source deviates from the usual trend of harder when brighter feature shown by blazars. To understand the likely physical scenario responsible for the flux variation, we performed a detailed broadband spectral analysis of the source by selecting different flux states from the multi-wavelength lightcurve. A single zone leptonic model was able to reproduce the broadband spectral energy distribution (SED) of each state. The parameters of the model in each flux state are determined using a $\chi^2$ fit. We observed that the synchrotron, synchrotron-self-Compton (SSC), and External-Compton (EC) processes produce the broadband SED under varied flux states. The adjoining contribution of the seed photons from the broad-line region (BLR) and the IR torus for the EC process are required to provide adequate fits to the GeV spectrum in all the chosen states.

\end{abstract}




\section{Introduction}
\label{sec:int}

The active galactic nucleus (AGN) is the central luminous region of galaxy which outshines the entire host galaxy.  The source of this enormous amount of energy is considered to be a super massive black hole, which accretes the surrounding matter and creates a disk-like structure known as an accretion disc. Blazars are a type of AGN's in which relativistic jet of matter is oriented nearer to the observer's line of sight \citep{1995PASP..107..803U}. This configuration leads to Doppler-boosted emission, giving rise to extreme features such as fast flux variability \citep{2003ApJ...596..847B} and continuum emission across the electromagnetic spectrum \citep{1997ARA&A..35..445U}. Blazar variability is well known, and it ranges from timescales of minutes \citep{2007ApJ...664L..71A} to years \citep{2013MNRAS.436.1530R}. Blazars are further classified as Flat spectrum radio quasars (FSRQ's) and  BL Lacertae (BL Lacs). The FSRQ spectra shows strong emission line features in the optical spectrum, whereas the emission line features are absent/weak in  optical spectra of BL Lacs \citep{1995PASP..107..803U}.

The SED of blazars is characterised by two broad humps, the peak of  low energy hump occurs in the optical/UV/X-ray band, while the high energy hump peaks in the $\gamma$-ray energies. The low energy hump is due to the synchrotron emission of relativistic electrons in the jet \citep{1978bllo.conf..328B, 1989MNRAS.241P..43G}, whereas the scenarios responsible for the high energy component are still under debate. The common explanation for the emission  of high energy hump is the inverse Compton (IC) scattering of low energy photons, the source of these photons can be the synchrotron photons referred as synchrotron self Compton  \cite[SSC, see e.g.,][]{1985A&A...146..204G, 2002A&A...384...56C} or  photons entering external to the jet known as external Compton  \citep[EC see e.g.,][]{1993ApJ...416..458D, 1994ApJ...421..153S, 2018MNRAS.477.4749C}. The other explanation used is the hadronic model in which proton-synchrotron emission or pion decay processes are believed to be responsible for the $\gamma$-ray emission from blazars \citep{1993A&A...269...67M, 2015A&A...573A...7W}.

4C\,+01.02 is an FSRQ source  located at a high redshift, $ z\sim2.099$ \citep{1995AJ....109.1498H} with coordinates, R.A = 17.1615 and  Dec = 1.58342 \citep{1995AJ....110..880J}. The source is also known by other names viz.  PKS\,0106+01 and OC\, 12 and has been studied previously by \citet{2011MNRAS.411..901G, 2017ApJ...851...33P, 2022ApJ...925..139S}. These studies mostly focus on the black hole mass estimations by modelling the SED of the source in different energy bands. In addition \citet{2022ApJ...925..139S} has used the thermal component from the accretion disk and non-thermal synchrotron emission in order to model the broadband observation and polarisation at optical energy in one of its brightest periods. The source has been reported in active state in various astronomical telegrams. For example, \citet{2013ATel.5394....1B} reported a daily $\gamma$-ray flux of $\rm (1.0 \pm 0.2) \times 10^{-6}\ ph\ cm^{-2}\ s^{-1}$ on 14 September 2013, this flux is $\sim 6$ times the average monthly integral flux (E $>$ 100 MeV) reported in the 2FGL catalog \citep{2012ApJS..199...31N}. The source was again reported in high flaring activity by \cite{2014ATel.6844....1C} on 20 December 2014 with a daily average $\gamma$-ray flux of $\rm (1.0 \pm 0.3) \times 10^{-6}\ ph\ cm^{-2}\ s^{-1}$.  It was again reported in high flux state on 20 November 2015 with a daily averaged $\gamma$-ray flux of $\rm (2.1 \pm 0.3) \times 10^{-6}\ ph\ cm^{-2}\ s^{-1}$  \citep{2015ATel.8319....1O},  which is $\sim 17$ times the average monthly integral flux (E $>$ 100 MeV) reported in the 3FGL catalog \citep{2015ApJS..218...23A}. The \emph{AGILE} detection of the source in increased flux state was reported in \cite{2016ATel.9232....1V} with an integrated $\gamma$-ray flux of $\rm (2.8 \pm 0.8) \times 10^{-6}\ ph\ cm^{-2}\ s^{-1}$. In the near infrared band the brightening of source was reported by \cite{2018ATel12243....1C} on 26 November 2018, in which the source increased its flux by a factor of 5. The enhanced activity at GeV for the source was again observed on 16 February, 2021 by \emph{Fermi}-LAT \citep{2021ATel14404....1H}. Moreover, \emph{Fermi}-LAT has observed the source in active state recently during 5-6 February, 2022 \citep{2022ATel15213....1M} and 11 March 2022 \citep{2022ATel15274....1P}. Despite such long list of high flux detection's, the detailed long term study of the source has not been carried out so far. In this work, we carried a detailed study of  the source by using  a long term multi-wavelength observations of the source from \emph{Fermi}-LAT, \emph{Swift}-XRT and \emph{Swift}-UVOT.
The structure of this paper is as follow: the multi-wavelength data details and the data analysis procedures are given in section \S \ref{sec:obs_data_ana},  the broadband spectral and  temporal results of  the source are presented in section \S \ref{sec:results}. Finally, results are summarised and discussed in section \S \ref{sec:discussion}. In this work we adopt a cosmology with $\rm \Omega_M = 0.3$, $\rm \Omega_\Lambda = 0.7$, and $\rm H_0 = 71  km s^{-1} Mpc^{-1}$.

\section{Observations and Data Analysis}
\label{sec:obs_data_ana}

\subsection{\emph{Fermi}-LAT}
\emph{Fermi}-LAT is one of the two instruments on board Fermi gamma-ray space telescope. It is pair conversion telescope capable of detecting high energy $\gamma$-rays in the energy range of 20 MeV - 1 TeV \citep{2009ApJ...697.1071A}. It scans the entire sky in 3 hours and runs in all-sky scanning mode. This has resulted in the identification of a large number of $\gamma$-ray sources. 4C\,+01.02 is listed among the regularly monitored source-list of the  \emph{Fermi}-LAT and is continuously monitored by \emph{Fermi}-LAT since 4th August 2008, 15:43:36 UTC. In order to carry the long term study, we have  analysed more than 12 years (2008-08-04 to 2021-03-31, MJD 54682.6--59302.6) of \emph{Fermi}-LAT  PASS 8 data. We have chosen the region of interest (ROI) as a 15\degree circular region centered at the 4C\,+01.02 location.  The photons collected in the energy range of 0.1 - 300 GeV are considered in the analysis and the latest version of \emph{fermipy} --v1.0.1 \citep{2017ICRC...35..824W} and \emph{fermitools}\footnote{\url{https://fermi.gsfc.nasa.gov/ssc/data/analysis/documentation/}} --v2.0.1 are used in the analysis. The latest instrument function (IRF) \emph{"P8R3 SOURCE V3"} has been used. To avoid the earth limb contamination, we have chosen the maximum zenith angle of 90\degree.  Based on the most recent \emph{Fermi}-LAT 4FGL catalog \citep{2020ApJS..247...33A}, we obtained XML model file that includes all sources in the ROI. While doing the analysis, the parameters of the sources within the ROI were set free and beyond this region, the parameters of the sources were freezed to their  4FGL catalog values. The Galactic diffuse emission model \emph{"gll iem v07.fits"} and an extragalactic isotropic emission component \emph{"iso P8R3 SOURCE V3 v1.txt"} are added to the model file during fitting to account for diffuse background emission. To assess the detection significance of the each source in the ROI, the test statistics (TS) defined as $\rm TS = 2 log \mathcal{L}$  is used, where $\mathcal{L}$ is likelihood parameter of the analysis \citep{1996ApJ...461..396M}. In this study, we have generated $\gamma$-ray lightcurves in two different time bins: 7-day, and 2-days, and also SED's for various activity periods.

\subsection{\emph{Swift}-XRT/UVOT}
Swift is a multi-wavelength space-based observatory carrying three instruments onboard: Burst Alert Telescope (BAT), X-Ray Telescope (XRT) and Ultraviolet/Optical Telescope (UVOT) \citep{2005SSRv..120..165B}. It observes the sky in several wavebands such as hard X-ray, soft X-ray, Ultraviolet and Optical. 4C\,+01.02 was observed by \emph{Swift} during quiescent as well as flaring states. However in this work, we consider the observations during the active phase of the source.
The complete details regarding the observations used from \emph{Swift}-XRT/UVOT are given in Table~\ref{tab:xrt_obs}.

For the XRT data, we have generated the cleaned event files corresponding to photon counting (PC) mode using \emph{"xrtpipeline"}, Version: 0.13.5 and  calibration file (CALDB, version: 20190910). The \emph{"xselect"} tool is used for selection of source and background regions, and the obtained spectrum files of the corresponding regions are saved. The source region is chosen from a  circular region of 25 pixels, while for the background, we have chosen a circular region of 50 pixels away from the source location.  Then \emph{"xrtmkarf"} tool has been used to generate the corresponding ancillary response file (ARF). In order to make the spectrum valid for the $\chi^2$ statistics, we used the \emph{"grppha"} task and binned the spectrum such that each bin contains minimum  of 20 counts. The X-ray spectral fit  in the energy range of 0.3-10 keV is done  by the \emph{"xspec"} tool \citep{1996ASPC..101...17A}, which is build in the \emph{HEASOFT} package. The absorbed powerlaw (PL)/broken powerlaw (BPL) model is used  to fit the spectrum. During the fit, the neutral hydrogen column density $\rm n_H$ is freezed to a value of $\rm 2.25 \times 10^{20}\ cm^{-2}$ \citep{2005A&A...440..775K}, while other parameters of the model are kept free.

The \emph{Swift}-UVOT telescope \citep{2005SSRv..120...95R} has also observed the FSRQ source 4C\,+01.02 along with the \emph{Swift}-XRT. It was monitored in all of the six filters viz U (3465 \AA), V (5468 \AA), B (4392 \AA), UVW1 (2600 \AA), UVM2 (2246 \AA) and UVW2 (1928 \AA). The \emph{"uvotsource"} task has been used to analyse the data acquired from the HEASARC archive. This task is build in the \emph{HEASOFT} (v6.27.2) package. The \emph{"uvotisum"} tool is used for adding the multiple images in a filter wherever present. Keeping the source at the centre of a circle, we extracted a source region of 5 arcsec around the source. The background region of larger area $\approx$3 times the source region has been selected. Following \cite{2011ApJ...737..103S}, the corresponding measured fluxes were de-reddened for Galactic extinction using $E(B - V) = 0.0208$ and $R_V = A_V /E(B - V) = 3.1$.

\begin{table*}
\centering
\begin{tabular}{l c c c c c}
\hline 
S. No. & Instrument & Observation ID & Time (MJD) & XRT exposure (ks) & UVOT exposure (ks)\\
\hline
\hline
1. & \emph{Swift}-XRT/UVOT & 00033561002 & 57014.5 & 4.93 & 1.62\\
2. & \emph{Swift}-XRT/UVOT & 00033561003 & 57016.4 & 5.03 & 1.65\\
3. & \emph{Swift}-XRT/UVOT & 00033561005 & 57357.3 & 3.69 & 1.22\\
4. & \emph{Swift}-XRT/UVOT & 00033561006 & 57360.6 & 3.87 & 1.27\\
5. & \emph{Swift}-XRT/UVOT & 00033561007 & 57363.5 & 3.94 & 1.29\\
6. & \emph{Swift}-XRT/UVOT & 00033561008 & 57728.9 & 2.77 & 0.62\\
7. & \emph{Swift}-XRT/UVOT & 00033561009 & 57967.9 & 1.41 & 1.41\\

\hline
\hline
\end{tabular}
\vspace{0.5cm}
\caption{List of \emph{Swift}-XRT/UVOT observations used in this work.\\ Note: The Observation ID 00033561009 has been skipped in broadband spectral analysis because of the lack of simultaneous multi-wavelength data in UV and optical bands.}
\label{tab:xrt_obs}
\end{table*}

\section{Results}
\label{sec:results}

\subsection{Temporal Study}
\label{subsec:temp_study}
\begin{figure*}
		\centering
		\includegraphics[scale=0.5]{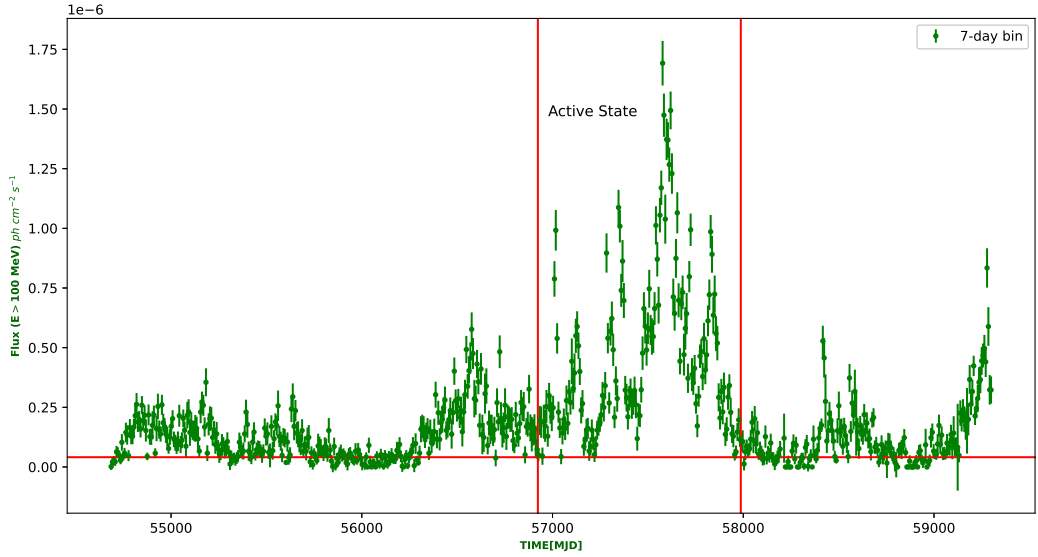}
		\vspace{0.5cm}
		\caption{7-day bin $\gamma$-ray lightcurve integrated over the energy range 0.1–300
GeV [(Flux (E$>$100 MeV)] of 4C\,+01.02 for the period 2008-08-04 to 2021-03-31(MJD 54682.6--59302.6). The time period from 2014-09-23 to 2017-08-24 (MJD 56923.1 -- 57989.1) between the red vertical lines is the active state of the source which is further analysed in this paper. The horizontal red line represents the base flux calculated
by averaging the flux points in the time interval, 2011-07-29 to
2012-11-16 (MJD 55771.1–56247.1), in which the source was in the
quiescent flux state.}
		\label{fig:7_day_fermi}
\end{figure*}

\begin{figure*}
		\centering
		\includegraphics[scale=0.5]{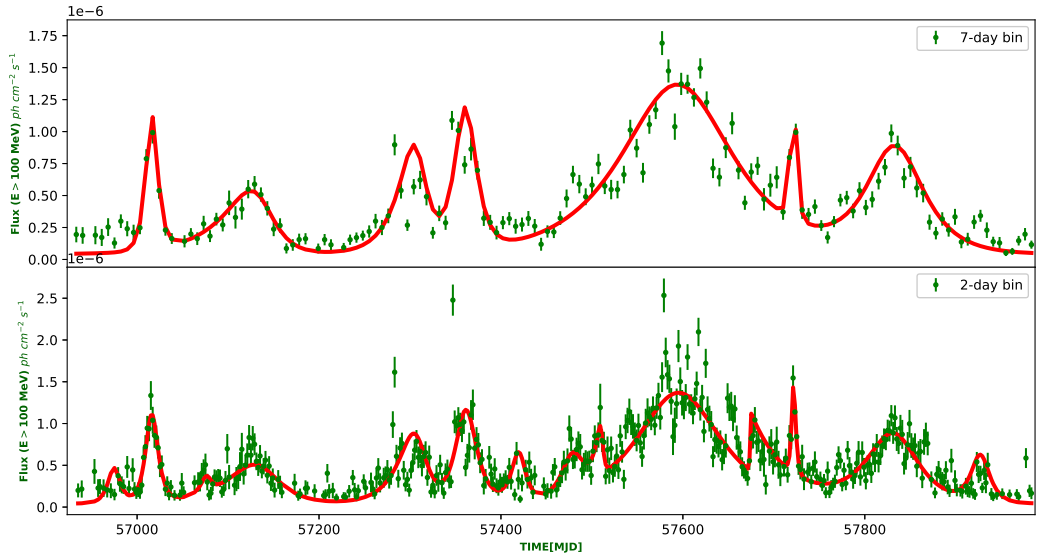}
		\vspace{0.5cm}
		\caption{Upper panel and lower panel represent the 7-day and 2-day bin $\gamma$-ray lightcurve integrated over the energy range 0.1–300 GeV [(Flux (E$>$100 MeV)] of 4C\,+01.02 during the active state 2014-09-23 to 2017-08-24 (MJD 56923.1 – 57989.1) respectively. The solid red curve in both panels is the best fitted sum of exponential functions.}
		\label{fig:7_2_day_fermi}
\end{figure*}

To study the temporal characteristics of the source, we obtained the long term $\gamma$-ray lightcurves of the source with time bin of seven and two days. In Figure~\ref{fig:7_day_fermi}, we have shown the seven day binned $\gamma$-ray lightcurve of the source integrated over the energy range 0.1--300 GeV, obtained during the time, 2008-08-04 to 2021-03-31 (MJD 54682.6--59302.6). We have excluded two weeks, 2018-03-19 to 2018-04-02 (MJD 58196.6--58210.6) data from the 7-day binned $\gamma$-ray lightcurve as there were no good time intervals (GTIs) available in this time interval. The horizontal red line in Figure~\ref{fig:7_day_fermi} represents the base flux of $\rm (4.1 \pm 0.3) \times 10^{-8}\ ph\ cm^{-2}\ s^{-1}$ calculated by averaging the flux points in the time interval, 2011-07-29 to 2012-11-16 (MJD 55771.1--56247.1), in which the source was in quiescent flux state. As shown in Figure \ref{fig:7_day_fermi}, the source displays significant variability across the light curve, with fractional variability amplitude of 1.14 $\pm$ 0.01. In order to carry detailed  temporal study of the source, we selected the time-interval 2014-09-23 to 2017-08-24 (MJD 56923.1--57989.1) during which the source was above the base line flux for nearly three years, we have named this time interval as the active state of the source. During the active state, the 7-day binned $\gamma$-ray light curve displays seven peak structures (see the top panel of Figure~\ref{fig:7_2_day_fermi}) with the maximum integral flux (E $>$ 100 MeV) of $\rm (1.7 \pm 0.1) \times 10^{-6}\ ph\ cm^{-2}\ s^{-1}$ obtained on 2016-07-08 (MJD 57577.1). Considering the good photon statistics in the active state, we obtained two day binned $\gamma$-ray light curve of the source in order to  acquire the adequate details of the variability.
The 2-day bin $\gamma$-ray lightcurve shows a maximum peak integral flux (E $>$ 100 MeV) of $\rm (2.5 \pm 0.2) \times 10^{-6}\ ph\ cm^{-2}\ s^{-1}$ on 2016-07-10 (MJD 57579.1), which is $\sim$ 61 times larger than the base flux. The 2-day bin $\gamma$-ray lightcurve in the active state shown in the bottom panel of Figure~\ref{fig:7_2_day_fermi} manifest seven additional components apart from the seven dominant components visible in the 7-day bin lightcurve. We estimated the rise and decay times of these components by fitting the exponential function \citep{2010ApJ...722..520A} to each of the component in the 2-day binned $\gamma$-ray lightcurve
 
\begin{equation}
    F(t) = F_b + \frac{2F_0}{\left[exp \left(\frac{t_0 - t}{T_r}\right) + exp \left(\frac{t - t_0}{T_d}\right)\right]}
    \label{eq1}
\end{equation}

where $F_b$ is the base flux, $t_0$ is the time corresponding to peak flux, $F_0$ of the component, $T_r$  and $T_d$ are the rise and decay time of the component. During the fit, we considered the flux points for which $\rm Flux/Flux_{err}>2$. The fitted profiles for the 7-day and 2-day binned lightcurve are shown in top and bottom panel of Figure~\ref{fig:7_2_day_fermi} respectively. Best fit parameter values for 2-day bin lightcurve are given in Table~\ref{tab:rise_decay}. Also we have measured the asymmetric parameter which gives the strength of symmetry of the components and is given by  $\zeta = \frac{T_d - T_r}{T_d + T_r}$. The component will be called as symmetric if the value of $\zeta < 0.3$, moderately asymmetric if $0.3 < \zeta < 0.7$ and asymmetric if $0.7 < \zeta < 1$ \citep{2010ApJ...722..520A}. The corresponding values of asymmetric parameter calculated for each component are given in Table~\ref{tab:rise_decay}. Looking at the values in table, components C1, C2, C3, C4, C6, C7, C8, C10, C12, C13 and C14  are symmetric, while C5 and C9 are moderately asymmetric and only C11 shows the  asymmetric behaviour. Further we calculated the shortest variability time by scanning the 2-day binned $\gamma$-ray lightcurve with the equation 

\begin{equation}
    F(t) = F(t_0) {2^{\frac{t-t_0}{\tau}}}
    \label{eq2}
\end{equation}

where $F(t_0)$ and $F(t)$ are the flux values measured at consecutive instants of time $t_0$ and $t$ respectively, $\tau$ represents the flux doubling or halving time. The values of these parameters are given in Table~\ref{tab:flux_doub_halv}. The positive or negative sign on the values of $\tau$ represents the flux doubling or halving time-scales, respectively. By applying the condition that the significance of difference in the flux at $t$ and $t_0$ is $\geq$ 3$\sigma$ \citep{2011A&A...530A..77F}, we found the shortest time variability of $0.66 \pm 0.08$ days between 2015-11-19 and 2015-11-21 (MJD 57345.1--57347.1) (see Table ~\ref{tab:flux_doub_halv}).

\begin{table*}
\centering
\begin{tabular}{c c c c c c} 
\hline 
Components & $\rm t_0$ & $\rm F_0$ & $\rm T_r$ &  $\rm T_d$ & $\rm |\zeta|$ \\
     &  (MJD) & ($\rm 10^{-6}\ ph\ cm^{-2}\ s^{-1}$) & (days) & (days) & \\
\hline
\hline
C1   &  56974.6 & 0.42 & 5.99 $\pm$ 0.91  & 4.0 $\pm$ 0.13 & 0.19\\
C2   &  57016.1 & 1.048 & 6.07 $\pm$ 0.35  & 4.97 $\pm$ 0.32 & 0.09\\
C3   & 57077.0 & 0.182 & 5.99 $\pm$ 3.46   & 4.00 $\pm$ 2.44 & 0.19\\
C4   &  57134.8 & 0.453 & 32.34 $\pm$ 3.62   & 19.99 $\pm$ 2.19 & 0.23\\
C5   &  57308.3 & 0.78 & 19.00 $\pm$ 0.99  & 9.68 $\pm$ 0.90 & 0.32\\
C6   &  57361.0 & 1.10 & 9.89 $\pm$ 0.66    & 10.96 $\pm$ 0.70 & 0.05\\
C7   &  57420.2 & 0.56 & 7.61 $\pm$ 0.89   & 6.03 $\pm$ 0.92 & 0.11\\
C8   &  57478.0 & 0.451 &  9.64 $\pm$ 1.40 & 10.95 $\pm$ 3.48 & 0.06\\
C9   &  57510.8 & 0.46 & 6.37 $\pm$ 1.99& 1.69 $\pm$ 0.70 & 0.57\\
C10   &  57594.8 & 1.32 & 40.71 $\pm$ 1.47  & 41.52 $\pm$ 1.34 & 0.009\\
C11   &  57673.2 & 0.36 & 0.06 $\pm$  0.04 & 39.52 $\pm$ 5.23 & 0.99\\
C12   &  57721.3 & 1.04 & 1.66 $\pm$  0.28 & 2.32 $\pm$ 0.31 & 0.16\\
C13   &  57832.4 & 0.82 &  26.81 $\pm$  1.68 & 23.69 $\pm$ 1.38 & 0.06\\
C14   &  57927.0 & 0.55 &  9.00 $\pm$  0.92 & 9.00 $\pm$ 0.86 & 4.59 $\times$ 10$^{-11}$\\
\hline
\hline
\end{tabular}
\vspace{0.5cm}
\caption{Rise and Decay times for the 2-day bin $\gamma$-ray lightcurve integrated over the energy range 0.1–300 GeV for the active state, 2014-09-23 to 2017-08-24 (MJD 56923.1 – 57989.1). Column description: 1. Individual Component, 2. Peak time (MJD), 3. Peak flux in units of $\rm 10^{-6}\ ph\ cm^{-2}\ s^{-1}$, 4. Rise time of the component, 5. Decay time of the component, 6. Asymmetry parameter.}
\label{tab:rise_decay} 
\end{table*}

\begin{table*}
\centering
\begin{tabular}{c c c c c c c} 
\hline 
$\rm T_{start}(t_0)$ & $\rm T_{stop}(t)$ & $\rm Flux_{start} [F(t{_0})]$ & $\rm Flux_{stop} [F(t)]$ &  $\rm \tau$ &  Significance & Rise/decay \\
(MJD) &  (MJD) & ($\rm 10^{-6}\ ph\ cm^{-2}\ s^{-1}$)& ($\rm 10^{-6}\ ph\ cm^{-2}\ s^{-1}$) & (days) & ($\rm \sigma$) & \\
\hline
\hline
57279.1 & 57281.1 & 0.37 $\pm$ 0.11 & 0.98 $\pm$ 0.15 & 1.44 $\pm$ 0.49 & 3.14 & R\\
57345.1 & 57347.1 & 0.30 $\pm$ 0.07  & 2.47 $\pm$ 0.18 & 0.66 $\pm$ 0.08 & 10.81 & R\\
57415.1 & 57417.1 & 0.14 $\pm$ 0.07 & 0.64 $\pm$ 0.13 & 0.95 $\pm$ 0.34 & 3.22 & R\\
57535.1 & 57537.1 & 0.33 $\pm$ 0.11 & 0.96 $\pm$ 0.14 & 1.30 $\pm$ 0.46 & 3.36 & R\\
57577.1 & 57595.1 & 1.55 $\pm$ 0.17 & 2.53 $\pm$ 0.20 & 2.84 $\pm$ 0.81 & 3.62 & R\\
57623.1 & 57625.1 & 0.90 $\pm$ 0.13 & 1.72 $\pm$ 0.17 & 2.16 $\pm$ 0.59 & 3.74 & R\\
57719.1 & 57721.1 & 0.81 $\pm$ 0.11 & 1.54 $\pm$ 0.15 & 2.16 $\pm$ 0.59 & 3.80 & R\\
57969.1 & 57977.1 & 0.12 $\pm$ 0.05 & 0.58 $\pm$ 0.11 & 3.56 $\pm$ 1.08 & 3.55 & R\\
\hline
\hline
\end{tabular} 
\vspace{0.5cm}
\caption{Flux doubling/halving time ($\rm \tau$) for each flare in the 2-day bin $\gamma$-ray lightcurve. R denotes the rise time of the component.}

\label{tab:flux_doub_halv} 
\end{table*}

\begin{figure*}
		\centering
		\includegraphics[scale=0.5]{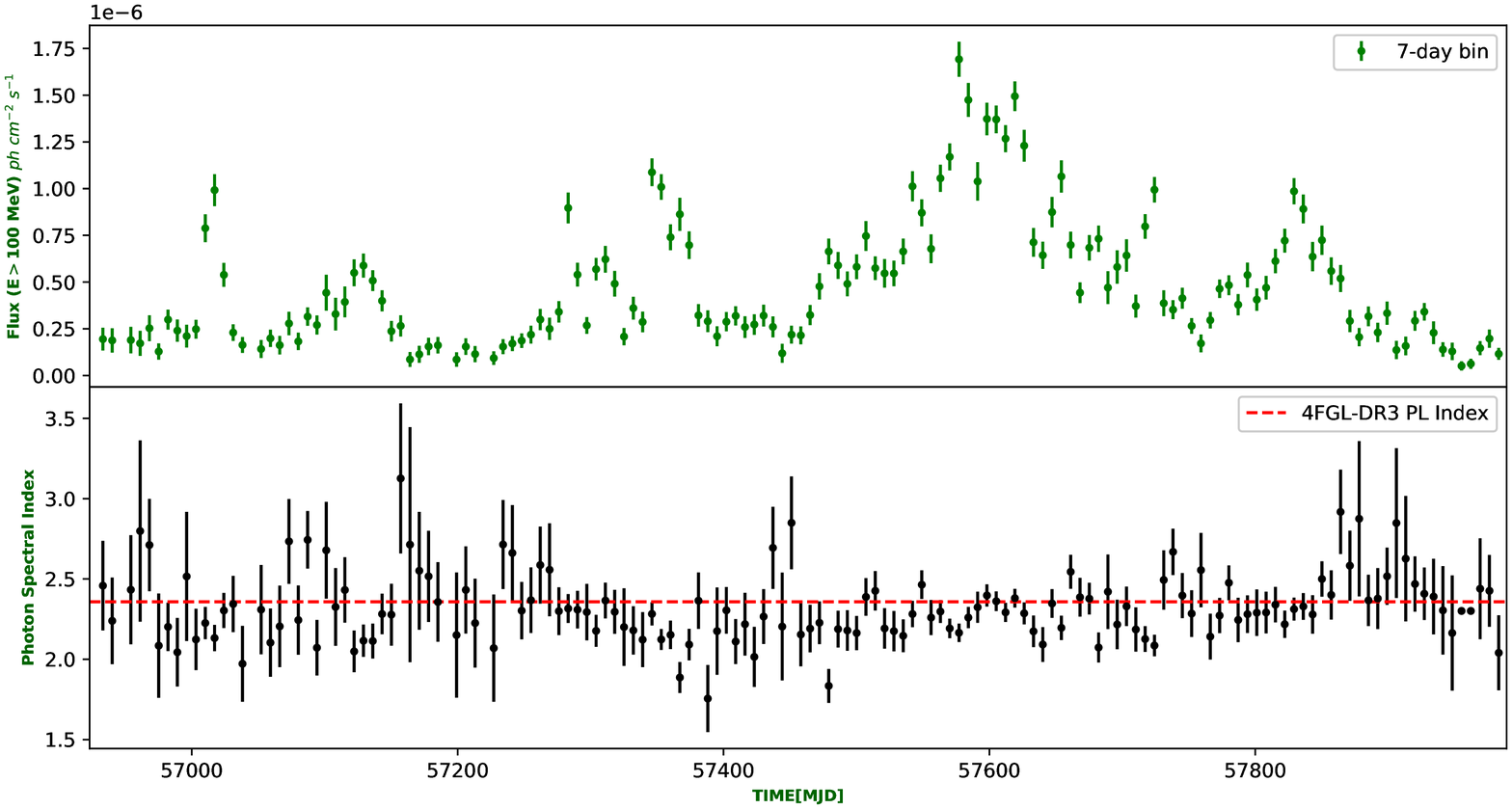}
		\vspace{0.5cm}
		\caption{Top panel is the 7-day bin $\gamma$-ray lightcurve of 4C\,+01.02 integrated over the energy range 0.1–300 GeV [(Flux (E$>$100 MeV)] for the active state, 2014-09-23 to 2017-08-24 (MJD 56923.1 – 57989.1), while bottom panel reports the corresponding values of modeled PL spectral index of the source. The red dashed line in the bottom panel represents the PL index reported in 4FGL catalog (Data Release 3).}
		\label{fig:flux_index}
\end{figure*}

\begin{figure*}
		\centering
		\includegraphics[scale=0.5]{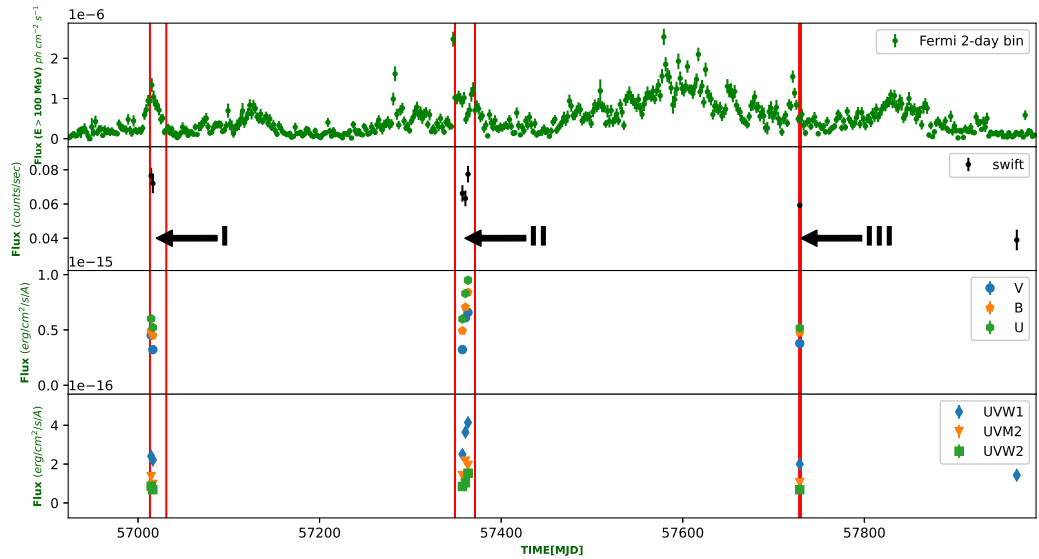}
		\vspace{0.5cm}
		\caption{Multi-wavelength lightcurve of 4C\,+01.02 by using the observations from \emph{Fermi}-LAT, \emph{Swift}-XRT and UVOT during the active state, 2014-09-23 to 2017-08-24 (MJD 56923.1 – 57989.1). Top panel represents the 2-day bin $\gamma$-ray lightcurve integrated over the energy range 0.1–300 GeV [(Flux (E$>$100 MeV)]. The second panel represents the X-ray lightcurve in the energy range of 0.3-10 keV. Third and fourth panel represents the optical and UV lightcurves. The regions included in the red vertical lines are flux states for which the broadband spectral behaviour of the source is studied.}
		\label{fig:mul_light}
\end{figure*}

During the active state of the source, the \emph{Swift}-XRT/UVOT carried a total of seven observations and the details of the observation are listed in Table~\ref{tab:xrt_obs}. We have used these observations to study the behaviour of the source at X-ray, UV and optical bands. The multi-wavelength lightcurve of 4C\,+01.02 during the active state using the observations from \emph{Fermi}-LAT, \emph{Swift}-XRT and UVOT is shown in Figure~\ref{fig:mul_light}. The upper panel of the multiplot represents the 2-day bin $\gamma$-ray lightcurve integrated over the energy range of 0.1--300 GeV, second panel represents the  X-ray lightcurve in the energy of 0.3-10 keV, and the optical/UV lightcurves are given in third and fourth panel respectively. 

In order to quantify the variability of the source with energy, we  calculated the fractional variability amplitude in different energy bands using the expression \citep{2003MNRAS.345.1271V} 

\begin{equation}
    \label{eq3}
    F_{var}=\sqrt{\frac{S^2-\overline{\sigma_{err}^2}}{\overline{F}^2}}
\end{equation}
where $\rm S^2$ is the variance, $\rm \overline{F}$ is the mean and $\rm \overline{\sigma_{err}^2}$ is the mean square of the measurement error on the flux points. The uncertainty on $\rm F_{var}$ is  given by \citep{2003MNRAS.345.1271V}

\begin{equation}
    \label{eq4}
    F_{var,err}=\sqrt{\frac{1}{2N}\left(\frac{\overline{\sigma_{err}^2}}{F_{var}\overline{F}^2}\right)^2+\frac{1}{N}\frac{\overline{\sigma_{err}^2}}{\overline{F}^2}}
\end{equation}

\begin{figure*}
		\centering
		\includegraphics[scale=0.5]{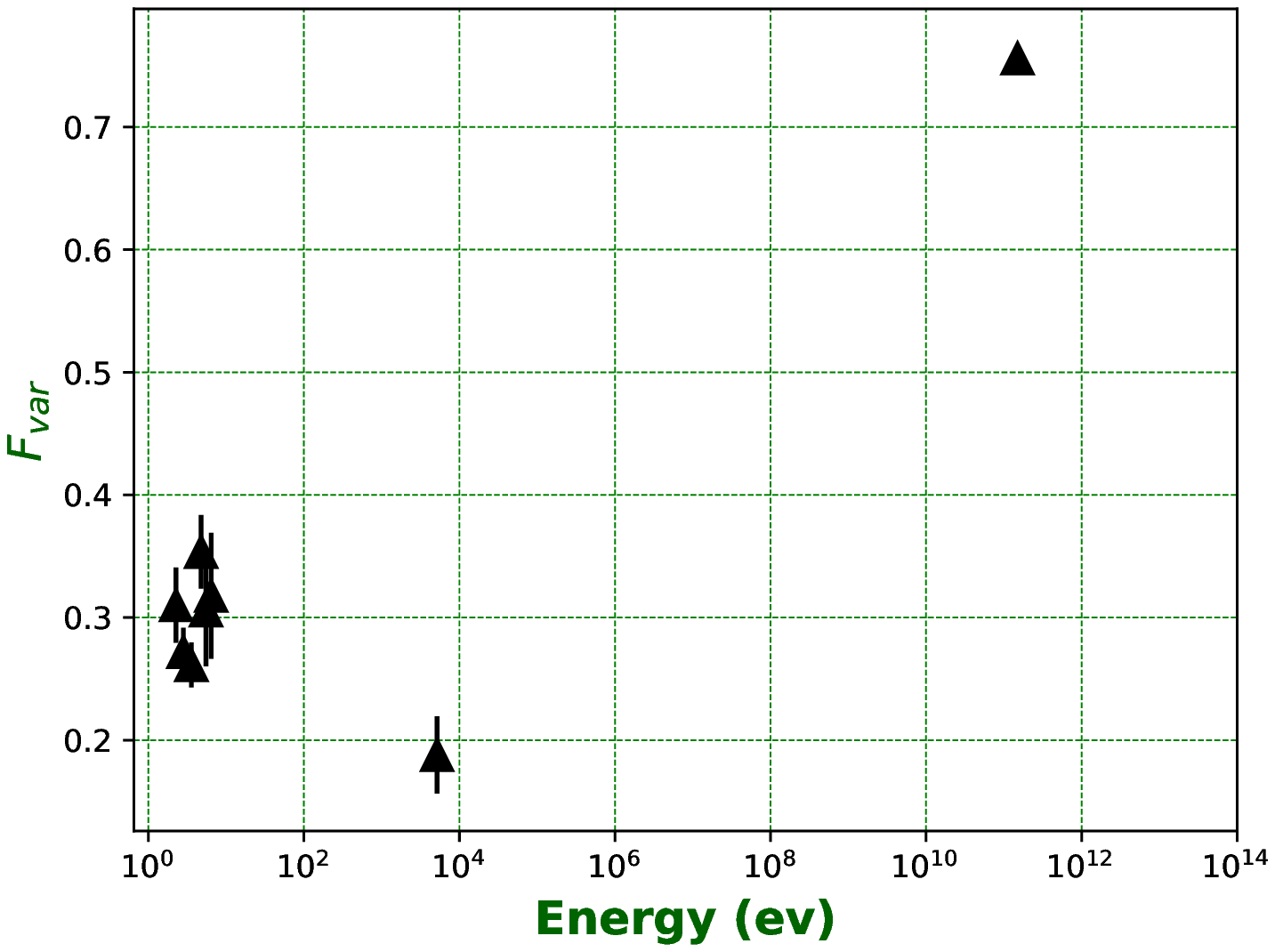}
		\vspace{0.5cm}
		\caption{Fractional variability amplitude ($\rm F_{var}$) obtained in different energy bands is plotted against the energy during the active state, 2014-09-23 to 2017-08-24 (MJD 56923.1 – 57989.1).}
		\label{fig:frac_var}
\end{figure*}

Here N represents the number of flux points in the lightcurve. The obtained values of $\rm F_{var}$ along with the uncertainties for different energy bands are given in Table~\ref{tab:frac_var} and the plot between $\rm F_{var}$ and energy is shown in Figure~\ref{fig:frac_var}. The corresponding $\rm F_{var}$ values shows a significant dip, while going towards the X-ray energy band and then the variability increasing in the GeV band. This dip in $\rm F_{var}$ vs energy plot is reported by several authors before \citep[see e.g.,][]{2016ApJ...819..156B, 2016A&A...590A..61C, 2017MNRAS.464..418R}. The dip can be related to the double hump feature observed in the broadband SED of blazars so that the electrons having higher energy dominate the emission at UV and GeV bands via synchrotron and IC process respectively,
while the X-ray  emission could be due to low energy tail of the particle distribution, which loses energy through IC process. As a result, the variability can be attributed to the cooling of relativistic electrons, such that the electrons with higher energy cool faster than the low energy electrons, giving rise to larger variability amplitude in case of emission from high energy electrons.

\begin{table*}
\centering
\begin{tabular}{l r}
\hline 
Energy band  & $\rm F_{var}$ \\
\hline
$\gamma$-ray (\mbox{0.1--300 GeV}) & 0.76$\pm$0.009  \\
X-ray (0.3--10 keV) & 0.19$\pm$0.03  \\
UVW2 & 0.32$\pm$0.05 \\
UVM2 & 0.31$\pm$0.04\\
UVW1 & 0.35$\pm$0.03\\
U & 0.26 $\pm$ 0.02\\
B & 0.27 $\pm$ 0.02\\
V & 0.31 $\pm$ 0.03\\
\hline
\end{tabular}
\vspace{0.5cm}
\caption{Fractional variability amplitude ($\rm F_{var}$) calculated in different energy bands for the source during the active state, 2014-09-23 to 2017-08-24 (MJD 56923.1 – 57989.1).}
\label{tab:frac_var}
\end{table*}

\subsection{$\gamma$-ray Spectral Behaviour}
\label{subsec:hard_brighter}
The 2-day bin $\gamma$-ray lightcurve in the active state is obtained by fitting a PL model in each time bin. The functional form of PL is defined as 

\begin{equation}
    \label{eq5}
    \frac{dN}{dE} = N_0 \left(\frac{E}{E_0}\right) ^{-\Gamma}
\end{equation}
where $\rm N_0$ is the normalisation, $\rm E_0$ is the pivot energy and $\Gamma$ is  the slope or PL spectral index.
The harder when brighter trend is the common characteristics observed in blazars \citep[see e.g.,][]{2016ApJ...830..162B, 2019MNRAS.484.3168S}.
 To examine this feature,  we checked the correlation between integrated $\gamma$-ray flux (E $>$ 100 MeV) and $\Gamma$. We obtained the Spearman rank correlation coefficient, $\rm r_s \sim0.06$ with a null hypothesis probability, $\rm P_{rs} \sim0.18$, which do not confirm the harder when brighter trend for 4C +01.02.
A possible reason for not obtaining the harder when brighter trend may be associated with the larger error on the spectral indices, as the  Spearman-rank correlation does not take the error on correlated quantities into account. In order to examine this bias, we checked  the correlation between the spectral index and integral flux (E $>$ 100 MeV) in the 7-day binned $\gamma$-ray lightcurve, where the errors on the indices are comparatively lower.
The correlation results are obtained as $\rm r_s \sim-0.18$ and $\rm P_{rs} \sim 0.02$, which again don't assert the harder when brighter behaviour in 4C\,+01.02. Additionally, we plotted the lightcurve of 7-day bin $\gamma$-ray flux (E $>$ 100 MeV) and corresponding PL spectral index in Figure~\ref{fig:flux_index}. We have also shown the average PL index of the source obtained from 4FGL catalog (Data release 3) \citep{2022arXiv220111184F} in the Figure~\ref{fig:flux_index}. Considering this average PL index as reference, the 7-day bin PL index does not show significant hardness within error bars during the high flux state. Hence, the results of Figure~\ref{fig:flux_index} further asserts that the  harder when brighter trend is missing in this source.

\subsection{Broadband Spectral Analysis}
\label{subsec:broadband_spec_analysis}
\begin{figure*}
    \centering
    \includegraphics[scale=0.4,angle=270]{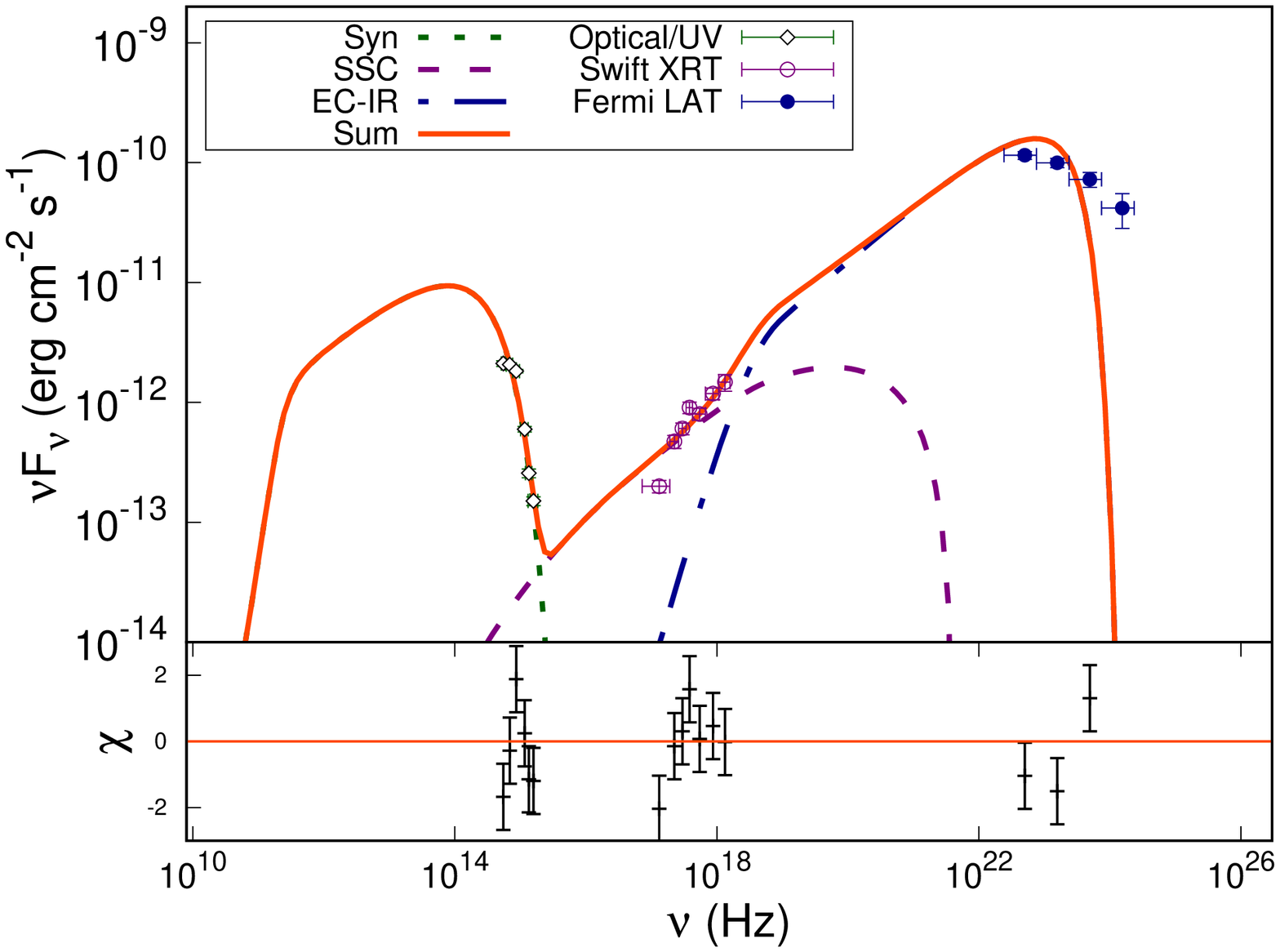}
    \includegraphics[scale=0.4,angle=270]{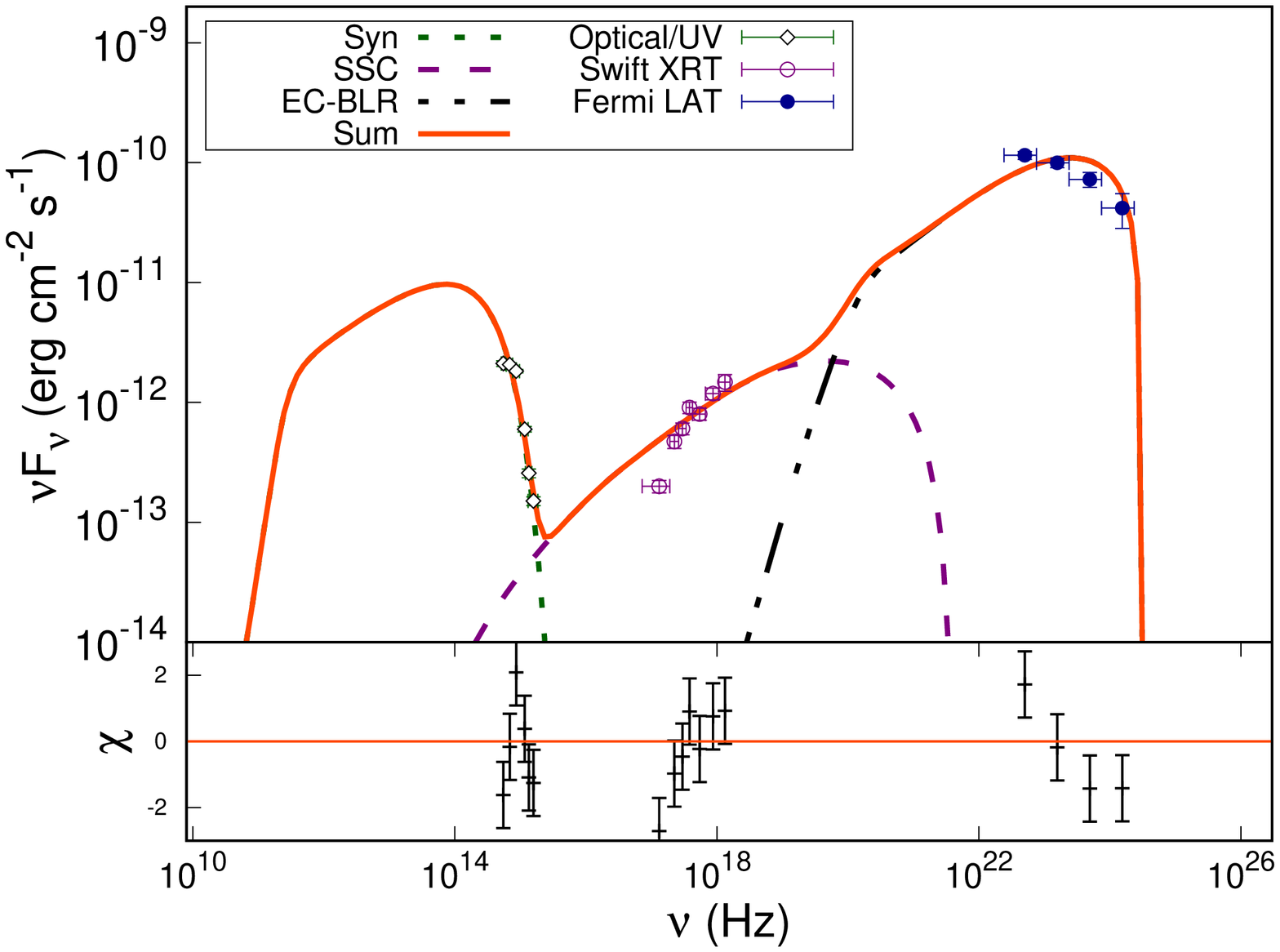}
    \includegraphics[scale=0.4,angle=270]{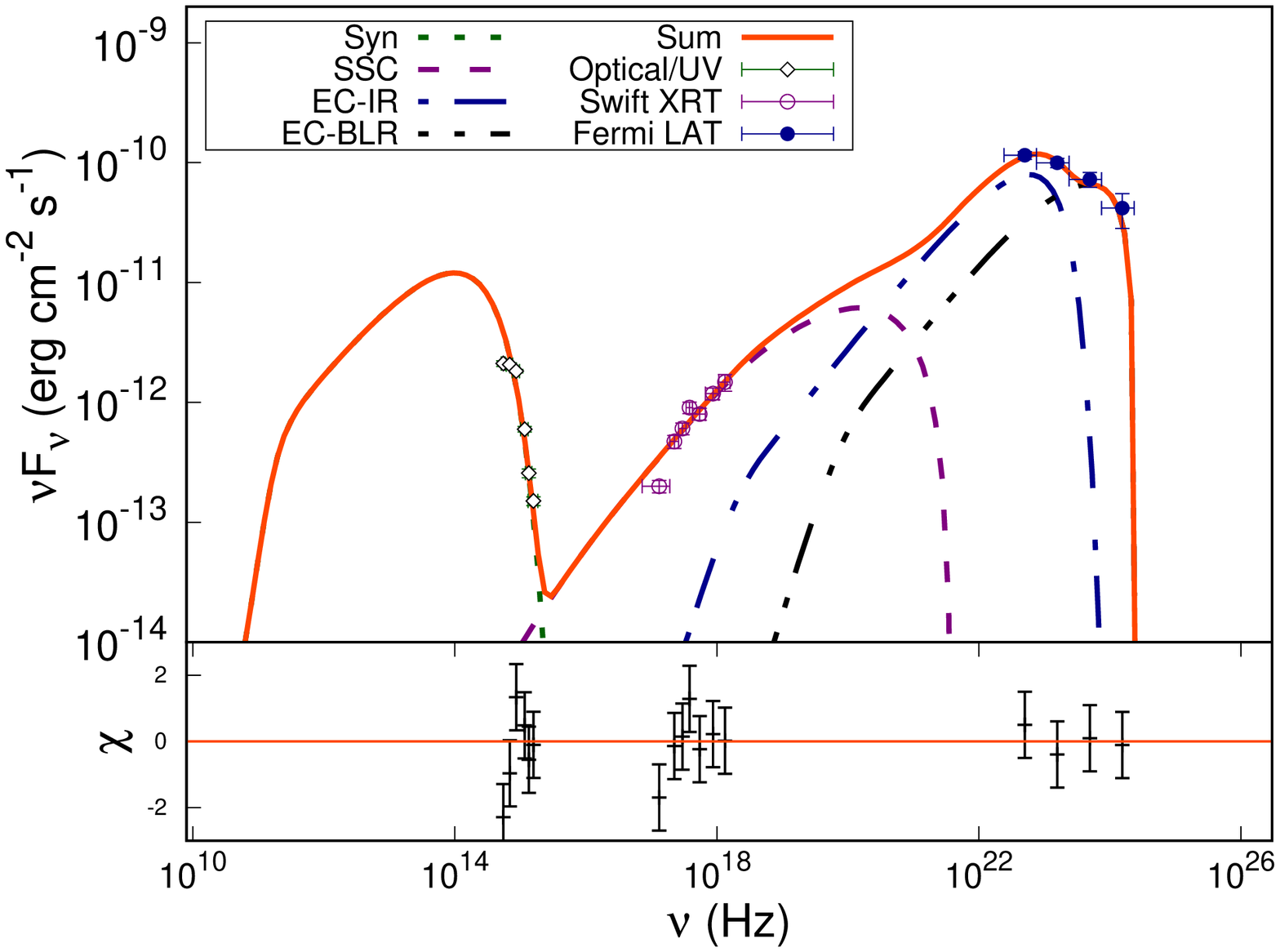}
    \vspace{0.5cm}
    \caption{Broadband SED of 4C\,+01.02 obtained during the I state. The observed flux points for corresponding energy bands are specified in different colors/symbols as mentioned in the labels. Upper left and upper right panels depict the seed photons responsible for the EC process as IR photons from the torus,  and BLR photons respectively,  while lower panel represents the combination of the two photon fields (EC/IR and EC/BLR).}
    \label{fig:bb_sed_I}
\end{figure*}

\begin{figure*}
    \centering
    \includegraphics[scale=0.4,angle=270]{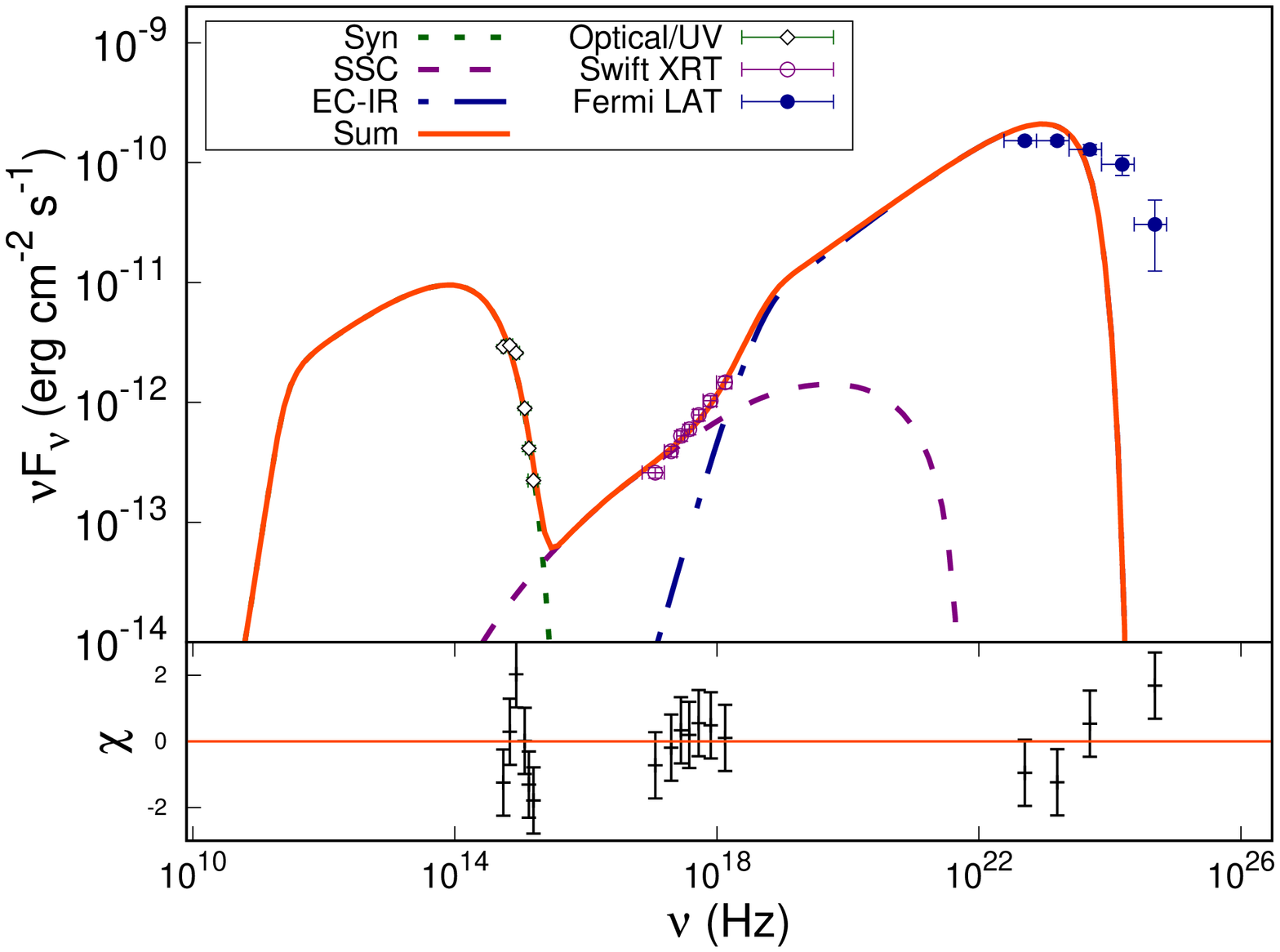}
    \includegraphics[scale=0.4,angle=270]{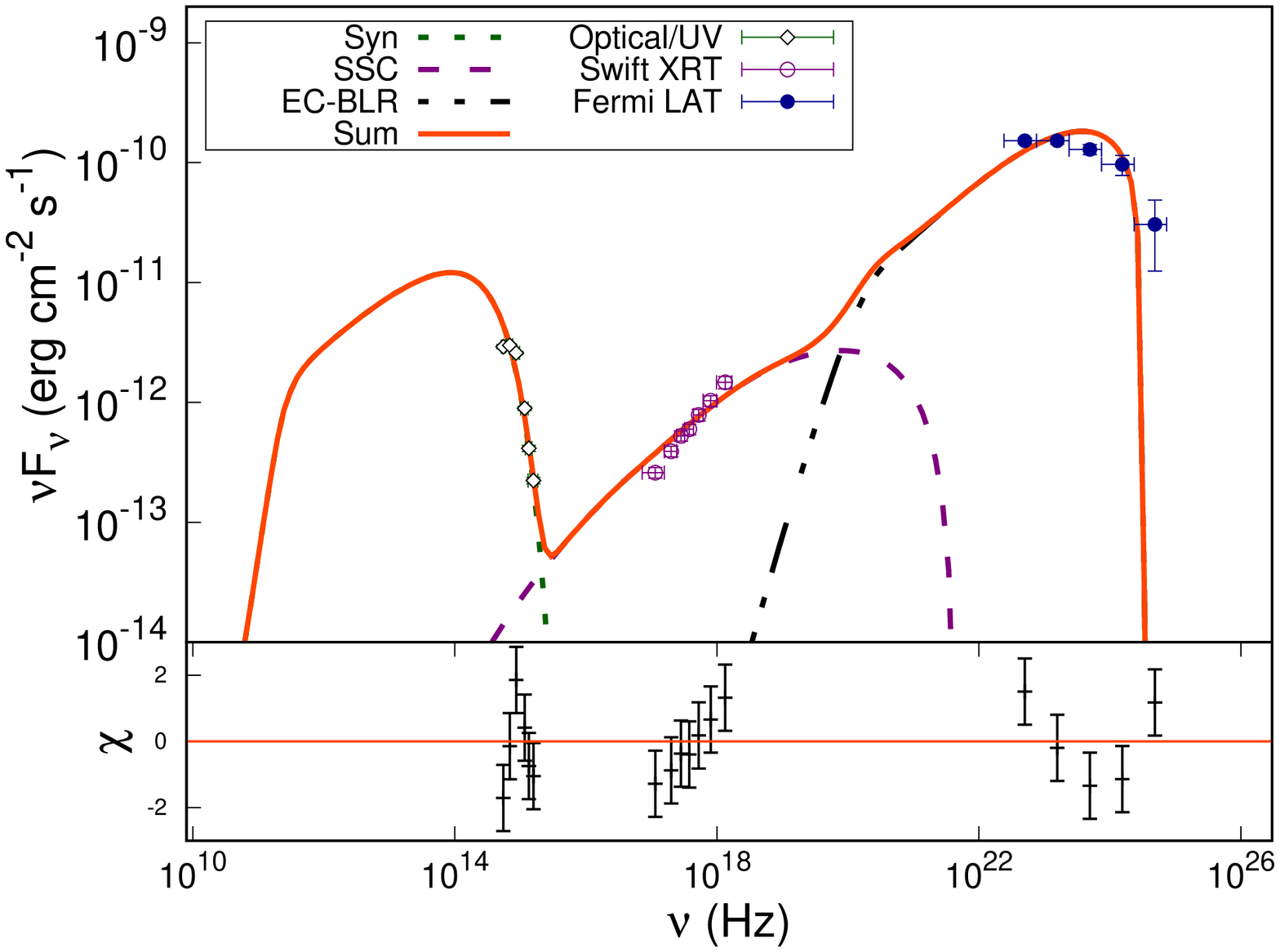}
    \includegraphics[scale=0.4,angle=270]{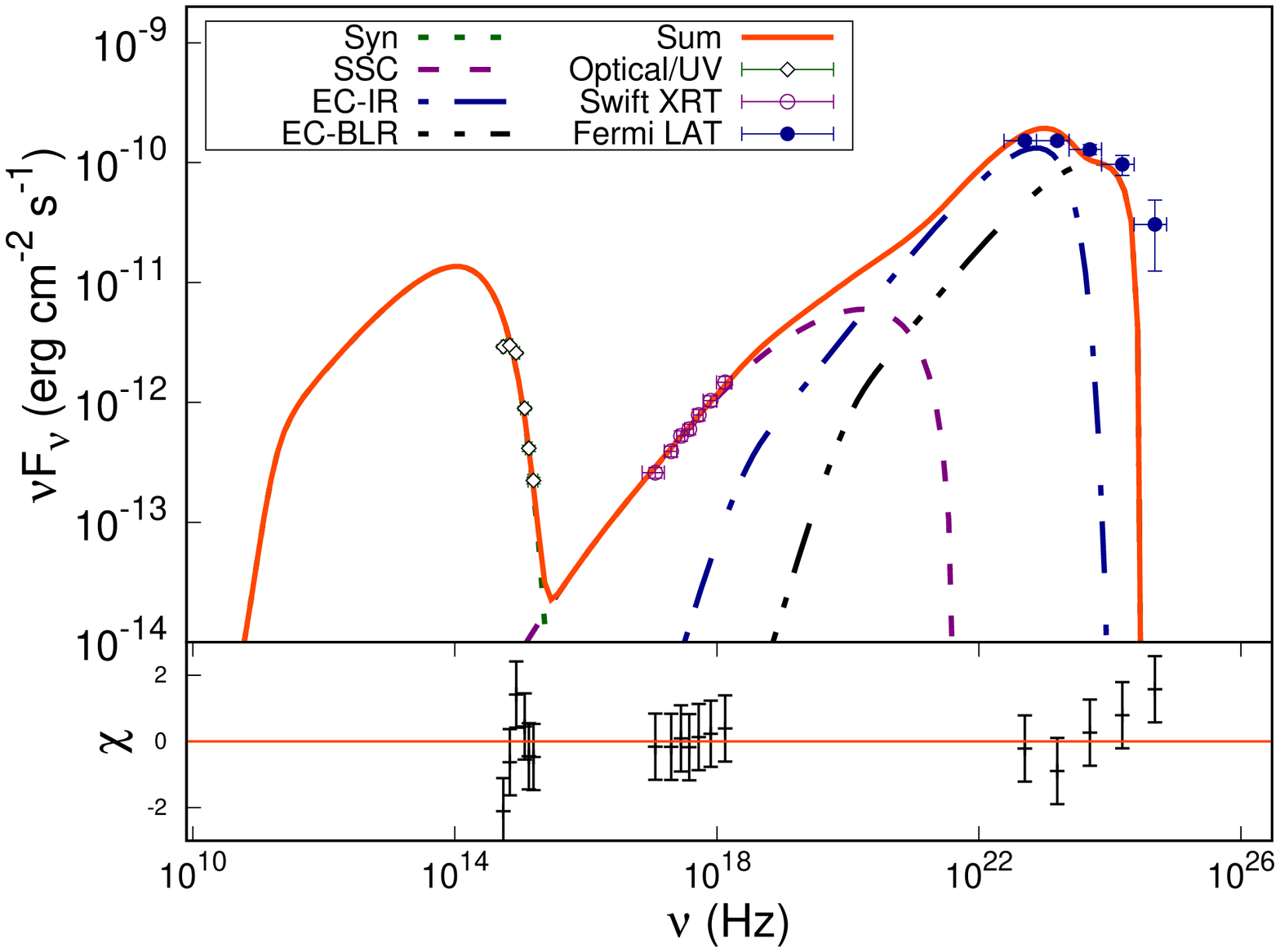}
    \vspace{0.5cm}
    \caption{Broadband SED of 4C\,+01.02 obtained during the II state. The labels are same as that of Figure~\ref{fig:bb_sed_I}}
    \label{fig:bb_sed_II}
\end{figure*}

\begin{figure*}
    \centering
    \includegraphics[scale=0.4,angle=270]{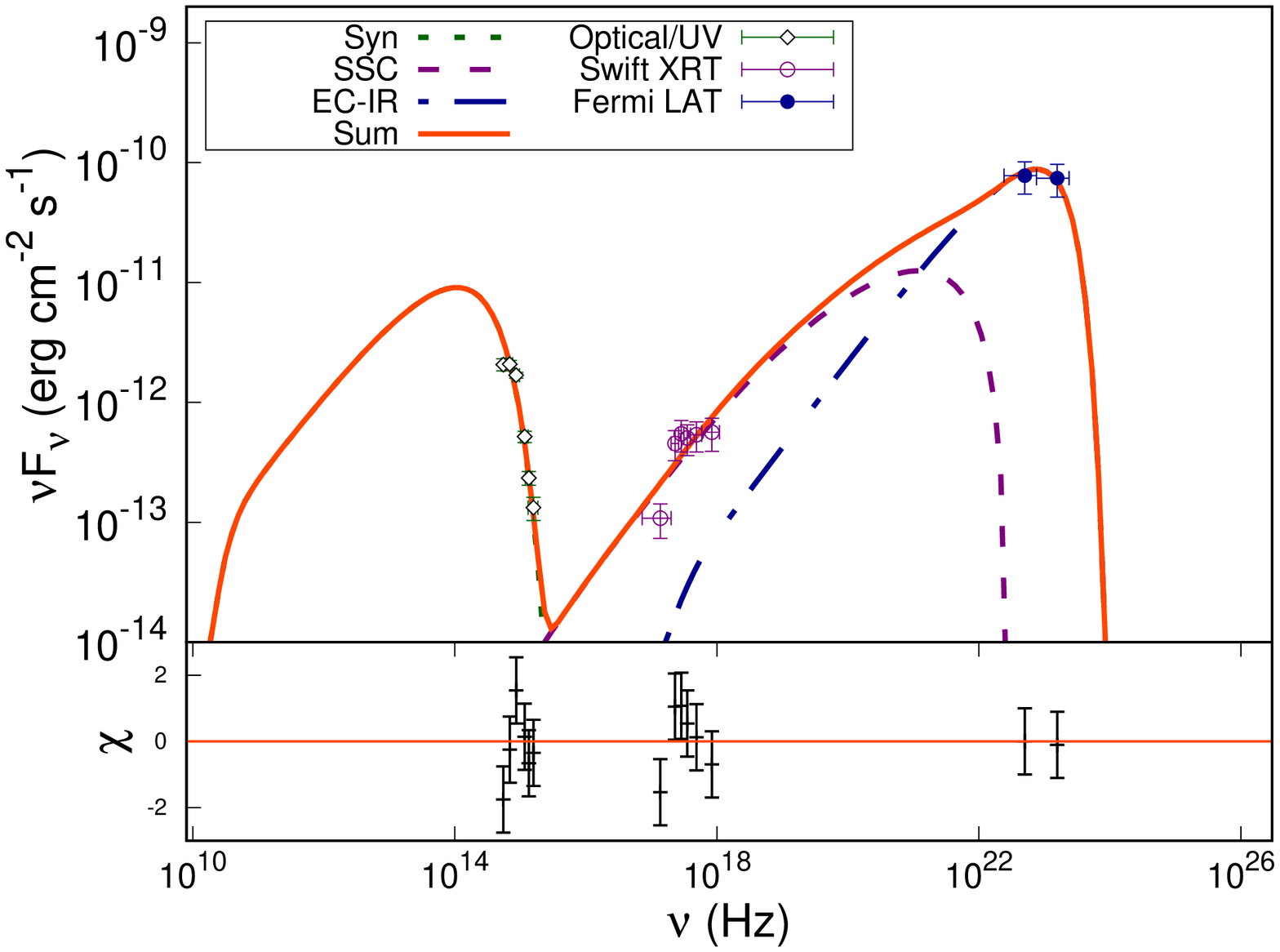}
    \includegraphics[scale=0.4,angle=270]{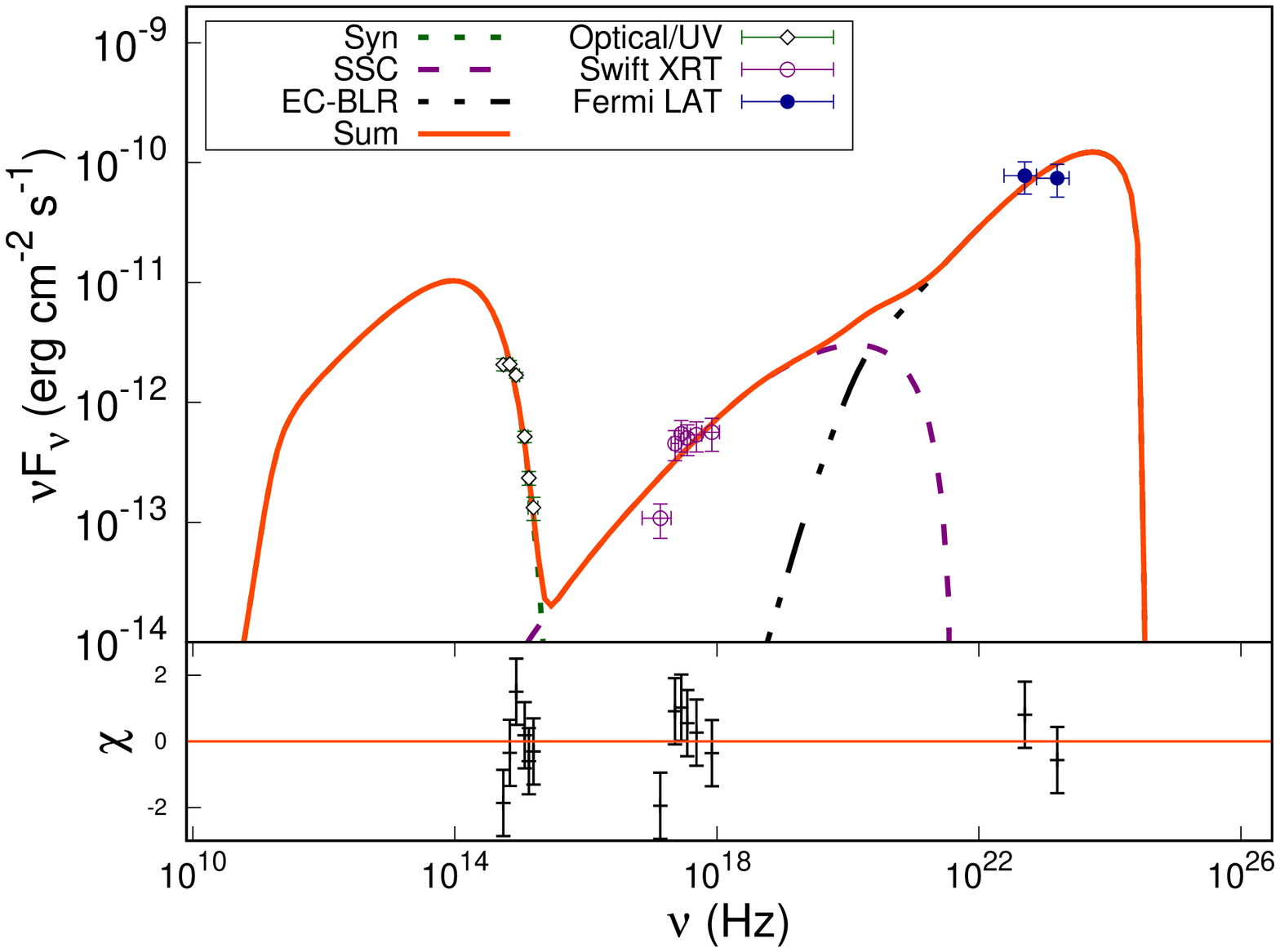}
    \includegraphics[scale=0.4,angle=270]{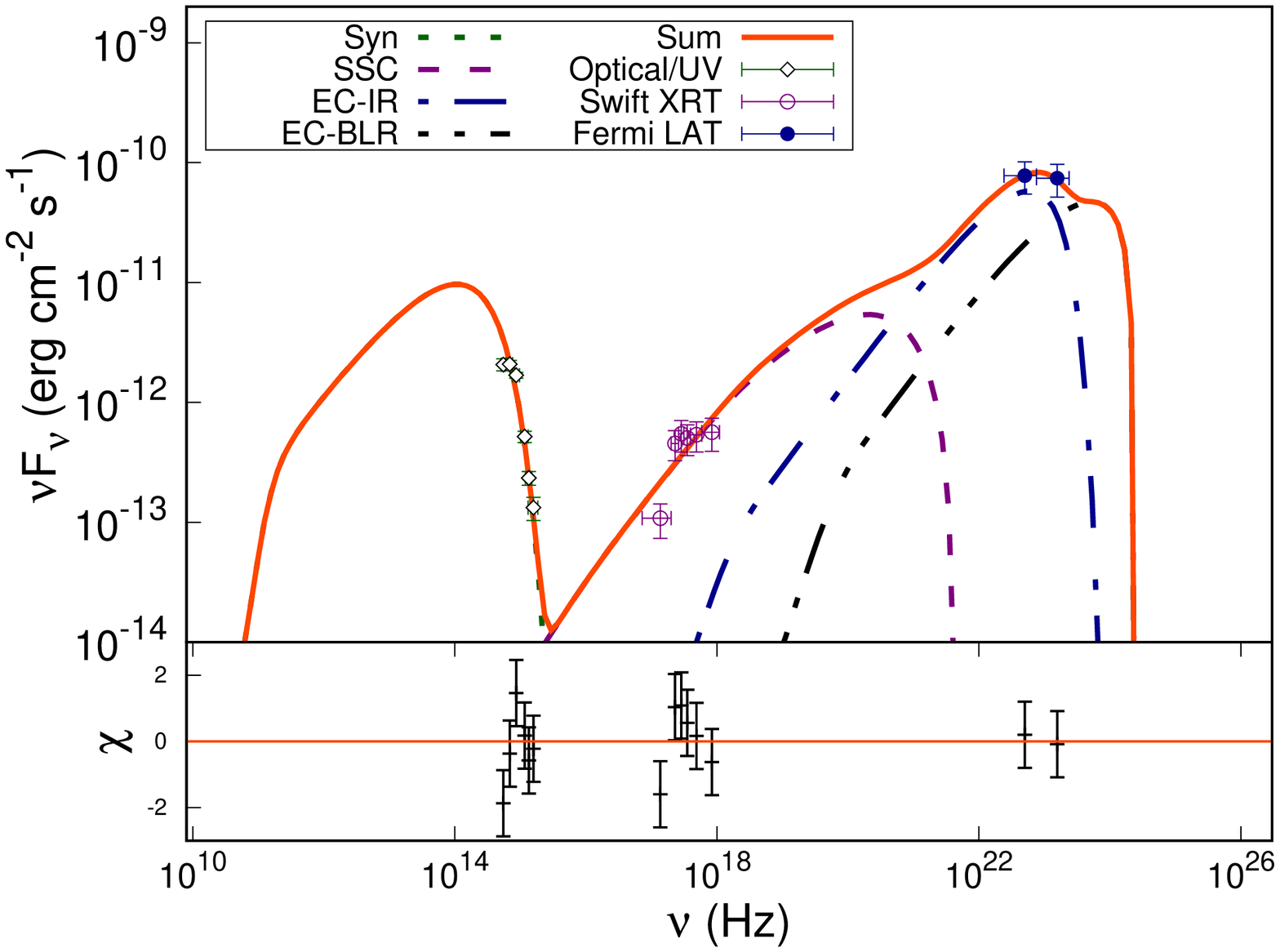}
    \vspace{0.5cm}
    \caption{Broadband SED of 4C\,+01.02 obtained during the III state.The labels are same as that of Figure~\ref{fig:bb_sed_I}}
    \label{fig:bb_sed_III}
\end{figure*}

In this section, we study in detail the broadband spectral characteristics of  4C\,+01.02 by using  the data from \emph{Fermi}-LAT,  \emph{Swift}-XRT and \emph{Swift}-UVOT.  The \emph{Swift}-XRT/UVOT carried a total of seven observations in the active state of 4C\,+01.02.  Six of the seven observations viz. 00033561002, 00033561003, 00033561005, 00033561006, 00033561007 and 00033561008 contains observation 
in the X-ray and all filters of the UVOT, whereas UVOT observation are missing in most of the filters in observation id 00033561009 (see Multiplot Figure~\ref{fig:mul_light}). The time intervals  for the broadband spectral study are chosen  based on the availability of simultaneous observations in Optical, UV, X-ray and $\gamma$-ray energies.
Therefore, we skipped the time containing the observation id 00033561009 and selected three time periods for which multi-wavelength data is available for the broadband spectral analysis . The corresponding time periods are highlighted by red vertical lines and  these states are denoted as I, II, III (from left to right see Figure~\ref{fig:mul_light}).  The $\gamma$-ray spectrum in the selected states is fitted with the PL model (defined in eq. \ref{eq5} ) and logparabola (LP) model. The functional form of the LP model is defined as 


\begin{equation}
    \label{eq6}
        \frac{dN}{dE} = N_0 \left(\frac{E}{E_0}\right) ^{-(\alpha + \beta \log(E/E_0))}
\end{equation}

where $\rm N_0$ is the normalisation, $\alpha$ is photon index at scale energy, $\rm E_0$ and  the parameter $\beta$ gives the measure of curvature in the spectrum. During the spectral fit, $\rm E_0$ is kept fixed at $\sim 499$ MeV, which is a default value in the \emph{Fermi} 4FGL catalog, while other parameters are kept free. 
The corresponding best fit parameters acquired in the three chosen states using PL and LP  models  are given in Table~\ref{tab:fermi_int_spec}. We calculated the test statistics of the curvature, which is defined as $\rm TS_{curve} = 2 [log \mathcal{L}(LP) - log \mathcal{L}(PL)]$ \citep{2012ApJS..199...31N} in order to assess the significance of the curvature in the $\gamma$-ray spectrum, $\rm TS_{curve} > 16$ implies a significant curvature. The obtained values of  $\rm TS_{curve}$ suggest a significant curvature for the states I and II while the curvature is insignificant in the spectrum of III state. Based on this we choose the $\gamma$-ray SED points for the I and II states from LP fit, while for III state, the SED points are selected from the PL fit. The SED points are obtained for each state by dividing the total energy, 0.1--300 GeV into 8 equally log spaced bins.  
In order to produce the X-ray spectrum, we used the \emph{"xselect"} tool to acquire the source and background files for each observation id. The corresponding  ancillary response function (ARF)  have been generated using the tool \emph{"xrtmkarf"}. \emph{"grppha"} task has been used to acquire 20 counts per bin. The grouped spectra is then fitted using the X-ray spectral-fitting package \emph{"xspec"} with PL and BPL model. The X-ray spectra of  I and II state is  better fitted with \emph{tbabs*PL} model, while \emph{tbabs*BPL} model provides better fit to the spectra of III state. In case of UVOT, We used the \emph{"uvotimsum"} tool to combine the images of each filter in particular state and then the flux values in each filter is obtained using the \emph{"uvotproduct"} task.
The broadband SED points for state I, II, III are shown in Figures~\ref{fig:bb_sed_I}, \ref{fig:bb_sed_II} and \ref{fig:bb_sed_III} respectively.
In order to model the broadband SED for the chosen states I, II and III,  we use one zone leptonic model involving synchrotron and IC processes \citep{2018RAA....18...35S, 2017MNRAS.470.3283S}. We assume that the emission arises from a spherical blob of radius R, which moves at relativistic velocity along the jet with bulk Lorentz factor $\rm \Gamma_b$ and at a small angle $\theta$ with respect to the observers line of sight. This relativistic motion at smaller angle causes the Doppler boosting of the observed flux, which is given by the beaming factor $\delta$. The emission region is assumed to  be populated with non-thermal relativistic electrons having a PL with an exponential cutoff energy distribution, defined as

\begin{align}
 \label{eq7}
 N(\gamma) d\gamma = K \gamma ^{-\alpha} exp {(-\gamma/\gamma_c)}, \qquad  \gamma > \gamma_{min}
\end{align}

Where K is the normalisation, $\gamma$ is the electron Lorentz factor, $\alpha$ is the PL index, $\rm \gamma_c$ and $\rm \gamma_{min}$ are the Lorentz factor corresponding to cut off and minimum energy of the electrons respectively. The relativistic electron distribution undergoes synchrotron losses in presence of tangled magnetic field, B and  IC losses in presence of low energy seed photons. The source of seed photons for IC process may be synchrotron photons (SSC process) or the photons entering external to the jet (EC process). Here we consider two cases of seed photons for EC process viz. IR photons from the dusty torus and Lyman-alpha line emission from the broad line region (BLR). The external target photon field is assumed to be blackbody with a temperature, $\rm T \sim 1000 K$ for IR photons and $\rm \sim 42000 K$ (which is equivalent to Lyman-alpha line emission) for BLR photons. We additionally enforce equipartition between magnetic field and particle energy density to limit the number of unbound parameters. The emissivities corresponding to synchrotron, SSC and EC process are solved numerically and the resultant numerical code is  incorporated as a local model in "\emph{XSPEC}" in order to fit the broadband SED of the source. The model related uncertainties and  systematic errors in the data are accounted by adding a systematic of 20\% to the data. The observed broadband SED is reproduced using the  following parameters $\rm \gamma_c$, $\alpha$,  B, R, $\rm \Gamma_b$, T, $\rm \gamma_{min}$, $f$ (fraction of photons which gets IC scattered) and $\eta$ (equipartiton parameter).
The limited information available at Optical/UV, X-ray, and $\gamma$-ray, prevents us to obtain unique set of model parameters. Therefore, to acquire the consistent solution of model parameters, we freezed the parameters  R, $\rm \gamma_{min}$, $\theta$ and $f$  and perform the broadband SED fit with $\alpha$, $\gamma_c$, $\rm \Gamma_b$ and B as free parameters. 
Since the  $\gamma$-ray SED points in the flaring state has better photon statistics, 
the typical values of the fixed parameters, $\rm R = 3.2 \times 10^{-16}cm$, $\theta = 1 \degree$ and $\rm \gamma_{min}$ = 10 are obtained by iteratively fitting the SED for I state. Also, the $f$ value for the IR and BLR photons are fixed to values $\sim 0.1$ and $\sim 10^{-8}$ respectively. The obtained fixed parameter values are then used to fit the broadband SED's for II and III states. The best-fit spectral model, which includes synchrotron, SSC, and EC components, as well as observed SED points for I, II and III states are shown in Figures~\ref{fig:bb_sed_I}, \ref{fig:bb_sed_II}, \ref{fig:bb_sed_III} and the corresponding fitting parameters are given in Table~\ref{tab:bb_parameters}. The model including the IR and BLR components individually does not yield a good fit to the broadband SED, instead we have to take both IR and BLR components into account in order to obtain a reasonable fit to the broad band SED in the specified flux states. In case of IR, BLR, and IR+BLR, the $\rm \chi^2/dof$ values are 30.79/13, 27.70/13, and 13.63/13 for state I, 41.65/14, 19.86/14, and 11.84/14 for state II, and 11.51/9, 13.13/10, and 11.74/10 for state III, respectively.

\begin{table*}
\centering
\begin{tabular}{c c c c c c c c c} 
\hline 
State & Period (MJD) & Model chosen & $\rm F_{ 0.1 - 300 GeV}$ & $\Gamma$ or $\alpha$ &  $\beta$ & $\rm TS_{value}$ & $\rm -log \mathcal{L}$ & $\rm TS_{curve}$\\
\hline
\hline
I & 57013.15 - 57031.15 & PL & 0.72 $\pm$ 0.06 & 2.29 $\pm$ 0.05  & --  & 777.04 & 6284.45 & -- \\
 &  & LP &  0.66 $\pm$ 0.06  & 2.21 $\pm$ 0.07  & 0.12 $\pm$ 0.05   & 784.15 & 6274.90 & 19.1  \\
II & 57349.15 - 57371.15 & PL &  0.99 $\pm$ 0.05  & 2.15 $\pm$ 0.03  & --  & 1966.27 & 8005.94 &  --\\
 &  & LP &  0.93 $\pm$ 0.05  & 2.07 $\pm$ 0.04 & 0.07 $\pm$ 0.02  & 2018.50 & 7995.59  &  20.4\\
III & 57728.15 - 57730.15 & PL & 0.48 $\pm$ 0.1   & 2.59 $\pm$ 0.24  & --  & 33.58  & 944.22 & -- \\
 &  & LP &  0.37 $\pm$ 0.1  & 3.10 $\pm$ 0.61  & 1.26 $\pm$ 0.72  & 43.65 & 940.60 & 7.24\\
\hline
\hline
\end{tabular} 
\vspace{0.5cm}
\caption{$\gamma$-ray integrated spectrum (E $>$ 100 MeV) for the chosen periods of time fitted with PL and LP models. Column description: 1. State chosen, 2. Time period of the state, 3. Corresponding model names, 4. Integrated flux in the units of $\rm 10^{-6}\ ph\ cm^{-2}\ s^{-1}$, 5. Corresponding photon index, 6. The curvature index, 7. Test statistics (TS), 8. -log(likelihood), 9. Curvature parameter.}

\label{tab:fermi_int_spec} 
\end{table*}


\begin{table*}
	\centering
\caption{Details of the fit parameters obtained by fitting the chosen states I, II, III using one zone SED emission model (\citealt{2018RAA....18...35S}). Column description; 1. Different flux states, 2. The location of the target photons in EC process, 3. Particle PL spectral index ($\alpha$) 4. Lorentz factor (\rm $\gamma_{c}$) at cut off energy, 5. Bulk Lorentz factor (\rm $\Gamma_b$), 6. Magnetic field (B), 7. ($\chi^{2}$/degrees of freedom) for a particular fit, 8. Doppler factor ($\delta$), 9. Jet power $(\rm P_{jet})$ in logarithmic scale $\rm (erg\ s^{-1})$, Subscripts and superscripts on the given values are the lower and upper error bounds respectively, while "--" indicates that upper or lower error bound is not constrained. Section \S \ref{sec:results} discusses and specifies the parameters that were kept frozen during the fitting of the spectrum of I, II and III states.} 
\vspace{0.5cm}
\begin{tabular}{c c c c c c c c c}
\hline
\hline
& & \multicolumn{5}{c}{Parameters} & \multicolumn{2}{c}{Properties} \\  \cmidrule(lr){3-7}  \cmidrule(lr){8-9}

 Flux & Seed & $\alpha$ & $\rm \gamma_{c}$ & $\rm \Gamma_b$ & B & $\chi^{2}$/dof &  $\delta$ & $\rm P_{jet}$ \\ 
state & photons & & & &  &  &  &  \\ \\
 
I & IR + BLR  &  $1.61_{-0.34}^{0.34}$ & $3111.09_{-208.01}^{189.65}$ & $11.95_{1.01}^{--}$ & $1.12_{-0.07}^{--}$ & 13.63/13 & 22.87 & 46.02  \\
\\

II & IR + BLR   &  $1.60_{-0.30}^{0.31}$ & $3195.75_{-220.56}^{229.36}$ & $13.51_{-1.11}^{--}$ & $1.03_{-0.06}^{--}$ & 11.84/14 & 25.58 & 46.04  \\
\\

III & IR + BLR   &  $1.50_{-0.37}^{0.33}$ & $3316.90_{-304.32}^{--}$ & $11.00_{-1.47}^{--}$ & $1.10_{-0.11}^{--}$ & 11.74/10 & 21.18 & 45.81 \\ 
\\
 
\hline 
\end{tabular}
\label{tab:bb_parameters}
\end{table*}

\section{Summary and Discussion}
\label{sec:discussion}

The long term monitoring of the FSRQ, 4C\,+01.02 by \emph{Fermi}-LAT and the availability of simultaneous observations by \emph{Swift}-XRT/UVOT has made it possible for the first time to perform a detailed  temporal and spectral study of the source. The 7-day bin $\gamma$-ray lightcurve of the source reveals a long period of flaring activity, we defined this period 2014-09-23 to 2017-08-24 (MJD 56923.1–57989.1) as the active state of the source. The two day bin $\gamma$-ray light curve shows a total of 14 flaring components in the active state (see Figure~\ref{fig:7_2_day_fermi}, bottom panel). While most of the components show symmetric behaviour, two of them are moderately asymmetric and one shows considerable asymmetric behaviour (\S \ref{sec:results}). 
 A significant asymmetric profile can be explained by fast injection of accelerated particles followed by delayed radiative cooling or escape from the emission region. 
 More asymmetric $\gamma$-ray flares are expected to be produced by short living fast electron injections and wider jet opening angles \citep{2010ApJ...722..520A}. On the other hand, symmetric flares  can be linked to the crossing time of radiation (or particles) across the emission zone, which is determined by geometry and spatial scales \citep{2000ApJ...542L.105T, 2001ApJ...563..569T}. The superposition and mixing of numerous short-duration events might likewise result in symmetric flare profile \citep{1999ApJS..120...95V}.

The spectral analysis of the $\gamma$-ray lightcurve in the chosen flux states shows a considerable curvature in I ($\rm TS_{curve} = 19.1$) and II ($\rm TS_{curve} = 20.4$) states. The common approach to understand the curvature in the spectra is to consider the radiative losses in the emission region \citep{2002MNRAS.336..721K}. However numerical calculations to solve the transfer equation of electrons and to introduce a curvature parameter makes this approach difficult. A different approach suggested by works like \citep{2004A&A...413..489M, 2018MNRAS.478L.105J, 2018MNRAS.480.2046G, 2021MNRAS.508.5921H} has shown that the particle acceleration/escape can lead to the significant curvature in the spectrum. \citet{2004A&A...413..489M} demonstrated that such curved features in particle distributions can arise when the acceleration probability is energy-dependent. Alternatively, the energy dependency of the escape timescale in the acceleration zone might be responsible for the spectral curvature \citep{2018MNRAS.478L.105J, 2018MNRAS.480.2046G, 2021MNRAS.508.5921H}. \citet{2021MNRAS.508.5921H} showed that the PL with maximum energy, energy dependent diffusion (EED) and energy dependent acceleration (EDA) models can produce the curvature in the spectrum. However, they rule out the PL with maximum energy model as the observed correlations were inconsistent with the predictions of the model. Also they have shown that, in case of EED and EDA, the observed correlations are in accordance with predictions of the model, but the parameters inferred from these models come out to be non physical. Therefore, to obtain better constrains on the physical scenarios responsible for the curvature in the spectrum, a more complex model like combination of EDD and EDA model needs to be taken into account.

The multi-wavelength temporal analysis of the source revealed a significant simultaneous variability in different energy bands which indicates that the electron distribution responsible for the emission is the same. The $F_{var}$ vs energy plot shows a low variability amplitude at low energies (X-rays and optical/UV bands) while it is large at $\gamma$-ray energies. This behaviour of the variability in different energy bands is consistent with previous results of blazars \citep[e.g, ][]{2005ApJ...629..686Z, 2019MNRAS.484.3168S}. The significant amplitude fluctuations at $\gamma$-rays imply that $\gamma$-ray emission is caused by high energy electrons, whereas small variations at low energy band may be caused by a low energy electron distribution. This is because high-energy electrons cool faster than the low-energy electrons. The rise in variability amplitude with increasing photon energy can also disclose the source's signature of spectral variability \citep{2005ApJ...629..686Z}.

The broadband SED for the chosen flux states is modelled using synchrotron, SSC and EC processes. The one zone leptonic model used here proves to be successful in modeling all the three flux states. In the previous studies (e.g., \citet{2019MNRAS.484.3168S, 2018RAA....18...35S}), BPL distribution of electrons has been used to model the broadband SED of blazars. However, the unusual 
behaviour of the source results in a much steeper spectrum at optical/UV band than the spectrum at GeV energies. We found that such broadband spectrum can not be modelled by using the BPL distribution of electrons. \citet{2022ApJ...925..139S} has modelled the SED of the source by considering the BPL with exponential cutoff distribution. In this work,  we show that the PL with exponential cutoff distribution can also model the SED satisfactorily, if one considers the contribution from both IR and BLR seed photons. The model gives the reduced $\chi^2$ nearly equal to one in all the flux states.
However, when the EC scattering of IR and BLR photons are considered separately, the fit obtained is poor, with reduced $\chi^2$ significantly larger than 1 (see Section \S \ref{sec:results}). The III state has just two $\gamma$-ray  data points in the broad band SED, it is impossible to constraint the model with such data statistics, thus based on the I and II state, we assume that EC scattering of IR and BLR photons is also responsible for the $\gamma$-ray emission in the III state. The reduced $\chi^2$ of $\rm \sim 11.74/10$ is obtained by considering the synchrotron, SSC and EC(IR+BLR) processes in the III state.
The best fit parameters obtained in the three flux states shows a variation in the magnetic field, its value ranges from 1.03--1.12 G. The  Doppler factor  shows an increasing trend from low flux state (III state) with $\delta\sim 21.18$ to the high flux state (II states) with $\delta\sim 25.58$. Although these values are higher than the average values for FSRQs \citep{2015MNRAS.448.1060G} but are within the physical realistic values \citep{2015MNRAS.454.1767L}. The estimated PL indices ($\alpha$) range from 1.50-1.60, which falls within the expected range of the standard particle acceleration models. For example, the spectra can be  very hard ($\sim 1$) to  steep  in the diffuse shock particle acceleration scenario, depending on shock speed, shock field obliquity, magnitude of turbulence and nature of particle scattering \citep{2012ApJ...745...63S}. The bulk Lorentz factor ($\rm \Gamma_b$) show a marginal change in different states, its value is lower in the III state and  increases slightly in other two states. The cut off energy, $\rm \gamma_c$ show an increasing trend from I to III state. We also calculated the jet power $\rm (P_{jet})$ using \citet{2008MNRAS.385..283C} by assuming that the number density of cold protons is equal to that of non thermal electrons. Table~\ref{tab:bb_parameters} displays the $\rm P_{jet}$ values for different states, which indicate a modest variance with the lowest value for the low flux state (III). Though our study shows a trend in the physical parameters from low flux state to high flux state, however, the errors on these parameters are not well constrained (see Table~\ref{tab:bb_parameters}). The reason is that we don't have the adequate optical/UV and X-ray observation available in the active state of the source. This constrained us to chose smaller intervals for the simultanuous broadband SED of the source, which resulted in  poor data statistics at the $\gamma$-ray band. Therefore, in the future work, more observations in optical/UV and X-ray energies are needed in order to obtain better constraint on the physical parameters of the source.

\section{Acknowledgement}
The authors thank the anonymous referee for valuable comments
and suggestions. M.Z, S.S, N.I \& A.M acknowledge the financial support provided by Department of Atomic energy (DAE), Board of Research in Nuclear Sciences (BRNS), Govt of India via Sanction Ref No.: 58/14/21/2019-BRNS. SZ is supported by the Department of Science and Technology, Govt. of India, under the INSPIRE Faculty grant (DST/INSPIRE/04/2020/002319). In this work, we have used the archival $\gamma$-ray data from Fermi Science Support Center (FSSC). We have also used the \emph{Swift}-XRT/UVOT data from the High Energy Astrophysics Science Archive Research Center (HEASARC), at NASA’s Goddard Space Flight Center.

\section{Data Availability}
All the data used here for analysis is publicly available and the results are incorporated in the article. The codes used in this work will be shared on the request to the corresponding author Malik Zahoor (email: malikzahoor313@gmail.com).

\bibliographystyle{mnras}
\bibliography{biblography} 


\bsp	
\label{lastpage}
\end{document}